\documentclass[manuscript]{acmart}

\usepackage{graphicx} 
\usepackage{todonotes}
\usepackage{caption}
\usepackage{subcaption}
\usepackage{xspace}
\usepackage{ framed}
\usepackage[utf8]{inputenc}
\usepackage{listings}

\usepackage{mdframed}
\usepackage{tikz}
\usepackage{algorithm}

\usepackage{algpseudocode}

\usetikzlibrary{positioning,calc}

\newcommand{\approach}{\textsc{Prompt2Constraints}\xspace}

\author{Ali Alfageeh}
\affiliation{
  \institution{University of Houston}
  \city{Houston}
  \state{Texas}
  \country{USA}
}
\email{aalfagee@cougarnet.uh.edu}

\author{Sadegh AlMahdi Kazemi Zarkouei}
\affiliation{
  \institution{University of Houston}
  \city{Houston}
  \state{Texas}
  \country{USA}
}
\email{skazemiz@cougarnet.uh.edu}

\author{Daye Nam}
\affiliation{
  \institution{University of California, Irvine}
  \city{Irvine}
  \state{California}
  \country{USA}
}
\email{dnam1@uci.edu}

\author{Daniel Prol}
\affiliation{
  \institution{Universidad Internacional de la Rioja}
  \city{Logroño}
  \country{Spain}
}
\email{daniel.prol403@comunidadunir.net}

\author{Matin Amoozadeh}
\affiliation{
  \institution{University of Houston}
  \city{Houston}
  \state{Texas}
  \country{USA}
}
\email{mamoozad@cougarnet.uh.edu}

\author{Souti Chattopadhyay}
\affiliation{
  \institution{University of Southern California}
  \city{Los Angeles}
  \state{California}
  \country{USA}
}
\email{schattop@usc.edu}

\author{James Prather}
\affiliation{
  \institution{Abilene Christian University}
  \city{Abilene}
  \state{Texas}
  \country{USA}
}
\email{james.prather@acu.edu}

\author{Paul Denny}
\affiliation{
  \institution{University of Auckland}
  \city{Auckland}
  \country{New Zealand}
}
\email{paul@cs.auckland.ac.nz}

\author{Juho Leinonen}
\affiliation{
  \institution{Aalto University}
  \city{Espoo}
  \country{Finland}
}
\email{juho.2.leinonen@aalto.fi}

\author{Michael Hilton}
\affiliation{
  \institution{Carnegie Mellon University}
  \city{Pittsburgh}
  \state{Pennsylvania}
  \country{USA}
}
\email{mhilton@cmu.edu}

\author{Sruti Srinivasa Ragavan}
\affiliation{
  \institution{Indian Institute of Technology, Kanpur}
  \city{Kanpur}
  \country{India}
}
\email{srutis@cse.iitk.ac.in}

\author{Mohammad Amin Alipour}
\affiliation{
  \institution{University of Houston}
  \city{Houston}
  \state{Texas}
  \country{USA}
}
\email{maalipou@central.uh.edu}

\begin{document}
\title{From Prompts to Propositions: A Logic-Based Lens on Student-LLM Interactions}
\renewcommand{\shortauthors}{Ali Alfageeh et al.}

\begin{abstract}

\textbf{Background and Context.}  The increasing integration of large language models (LLMs) in computing education presents an emerging challenge in understanding how students use LLMs and craft prompts to solve computational tasks.  Prior research has used both qualitative and quantitative methods to analyze prompting behavior, but these approaches lack scalability or fail to effectively capture the semantic evolution of prompts.

\noindent \textbf{Objective.}  In this paper, we investigate whether students' prompts can be systematically analyzed using propositional logic constraints.  We examine whether this approach can identify patterns in prompt evolution, detect struggling students, and provide insights into effective and ineffective strategies.

\noindent \textbf{Method.}  We introduce \approach, a novel method that translates students' prompts into logical constraints.  The constraints are able to represent the intent of the prompts in succinct and quantifiable ways.  We used this approach to analyze a dataset of 1,872 prompts from 203 students solving introductory programming tasks.

\noindent \textbf{Findings.} We find that while successful and unsuccessful attempts tend to use a similar number of constraints overall, when students fail, they often modify their prompts more significantly,  shifting problem-solving strategies midway.  We also identify points where specific interventions could be most helpful to students for refining their prompts. 

\noindent \textbf{Implications.} This work offers a new and scalable way to detect students who struggle in solving natural language programming tasks.  This work could be extended to investigate more complex tasks and integrated into programming tools to provide real-time support.

\end{abstract}


\maketitle

\section{Introduction}

With the increasing accuracy and usability of large language models (LLMs), there has been a remarkable surge in the popularity of LLMs among students. 
Students have reported using LLMs in a wide variety of applications relevant to their academic life~\cite{amoozadeh2024trust}, e.g., in writing and editing essays, generating graphics, brainstorming, etc.
Since programming is an important part of the computer science (CS) curriculum, CS students use LLMs to solve various programming activities, such as  debugging~\cite{10.1145/3641554.3701974}, code explanation~\cite{leinonen2023comparing}, and helping to solve programming assignments~\cite{denny2021promoting}.

Users interact with LLMs in a sequence of textual queries known as prompts. 
Prompts elaborate a user's intention through words.
The main differentiating factor between traditional information gathering, e.g., Google search, and LLM systems is that the conversational interaction modality affords an iterative process wherein users can refine and revise prompts based on the response from LLM. 
Due to the opacity of LLMs and their stochasticity, it is difficult to provide users with a mental model to predict the behavior of the LLMs with certainty, which makes general recommendations for good prompting a difficult task. 

Understanding how students develop and use prompts in their academic tasks is necessary for developing effective techniques and methodologies to encourage positive LLM uses or identify and deter maladaptive scenarios.
There have been several studies to understand students' prompting behavior in solving programming tasks, e.g.,~\cite{amoozadeh2024student,lau2023ban,prather2023s, denny2024computing}. 
They often include manual sensemaking of the prompts that would require extracting  the meaning of the prompt. 
Existing qualitative approaches often rely on manual sensemaking, which is both time-consuming and hard to scale to large datasets of student prompts. Quantitative techniques, on the other hand, tend to reduce prompts to mere word counts or keyword searches, overlooking the deeper semantic changes that occur during a multi-turn interaction.

In this paper, we introduce \approach, a systematic framework for translating each prompt into a set of logical constraints that capture its core requirements, such as language choices, function naming, parameter specifications, and output format. By viewing prompts through the lens of propositional logic, researchers and educators can directly compare how students evolve their problem-solving strategies: which constraints they add or remove, how they respond to failures, and the extent to which they backtrack or shift focus mid-session.  In doing so, \approach not only accelerates large-scale prompt analysis but also offers meaningful insights into where students may need targeted hints or interventions. Moreover, the framework’s ability to distill prompts into interpretable constraints allows for more robust benchmarking, facilitating comparisons of different tasks, cohorts, or LLM models.

We apply \approach to a dataset of $1872$ prompts collected from $203$ students working on three introductory Python problems. Our findings reveal distinct patterns of prompt evolution among successful and unsuccessful learners, illuminate how and when learners refine or abandon constraints, and highlight potential opportunities for early detection of cognitive struggles. These insights point to the transformative potential of \approach in the development of intelligent tutoring systems, the improvement of prompt-based pedagogy, and the advance of our greater understanding of human-AI interactions in computing education.

\paragraph{Contributions.}
The main contributions of this paper is twofold. 
\begin{itemize}
    \item It proposes an automated technique for mapping prompts to logical constraints.
    \item It uses this technique to analyze student prompting behavior in solving three programming problems. 
\end{itemize}

In this paper, we seek to answer the following research questions. 
\begin{itemize}
    \item \textbf{\textit{RQ1:}} How robust and accurate is \approach for automatically extracting constraints from prompts?
    \item \textbf{\textit{RQ2:}} How do students’ prompts evolve when solving programming tasks?
    \item \textbf{\textit{RQ3:}} How do successful and unsuccessful prompting strategies differ?
\end{itemize}

\section{Motivating Example}
Table~\ref{tab:motivating_example} shows an example of the application of this approach on a sequence of prompts in a student interaction to solve a programming task in the ~\cite{denny2024prompt} dataset. It shows a high-level view of the student's problem solving steps that can be used in other analyses. 
Each row of Table~\ref{tab:motivating_example} contains the prompt's text, and \approach-generated corresponding propositional logic constraints and justification for the changes in the constraints compared to the previous prompt.
\approach also generates the details of constraints that we removed for brevity.
The first prompt ``Write me a Python function that returns how many elements in a given list is the integer 0 '' into three constraints: 
$C1$ aims to constrain the LLM response to Python functions, 
and $C2$ and $C3$ aim to constrain the behavior of the function to producing the number of elements in the list ($C2$), that are zero ($C3$).
In the subsequent prompt, P2, the student adds a new constraint about the signature of the function ($C4$). 
Next, in P3, the student modifies the parameter name in the signature, where \approach successfully captures it, modifying the constraint $C4$ to $C5$.

\begin{table*}
    \centering
    \caption{Motivating Example}
    \begin{tabular}{|c|p{5cm}|c|p{5cm}|} \hline
         Prompt\# & Prompt text  & Translation to constraints& \approach-generated justification for the changes\\ \hline
        P1 & Write me a Python function that returns how many elements in a given list is the integer 0 & $C1 \land C2 \land C3$& \\ \hline
        P2 & Write me a Python function called counter(test\_input) that returns how many elements in a given list is the integer 0 & $C1 \land C4 \land C2 \land C3$ & $C4$: The function must now be named counter and take a parameter named test\_input.\\ \hline
        P3 & Write me a Python function called counter(user\_input) that returns the amount of times a element in a given list is the integer 0 & $C1 \land C5 \land C2 \land C3$& 
        C5 evolves from having a parameter named test\_input to user\_input. 
        \\ \hline
        \end{tabular}
    
    \label{tab:motivating_example}
\end{table*}

\section{Related Work}
\label{sec:related}
\subsection{LLMs for code generation and programming tasks}
Recent studies have explored how LLMs like GitHub Copilot and ChatGPT support programming. Denny et al.~\cite{denny2023conversing} evaluated Copilot’s performance on 166 CodeCheck exercises, finding that while 79 (47.6\%) of problems were solved on the first attempt, 53 (31.9\%) required modified prompts, and 34 (20.5\%) remained unsolved. This highlights both the potential and limitations of LLMs as programming assistants. Beyond code generation, Nam et al.\cite{nam2024using} investigated how LLMs aid code comprehension. Their IDE plugin, GILT, allows users to query LLMs for explanations, API details, and usage examples without explicit prompts. A user study showed significant improvements in task completion rates compared to web search, though benefits varied by experience level. Similarly, Etsenake and Nagappan\cite{etsenake2024understanding} analyzed human-LLM interactions and found that LLMs boost productivity but with mixed results depending on task complexity and user expertise. To enhance programming education, Denny et al.\cite{denny2024explaining} explored using LLMs to evaluate students' natural language explanations of code, reinforcing both comprehension and prompt engineering skills. In another study, they introduced "Prompt Problems"\cite{denny2024prompt}, a novel exercise type where students craft prompts to generate functional code, using a web-based tool called Promptly. Evaluations showed that students found this method engaging and valuable for developing computational thinking. Taking a different approach, Lane and VanLehn~\cite{lane2003coached} proposed coached program planning, a dialogue-based tutoring method guiding novice programmers through natural language pseudocode before coding. Their study found that students using this approach wrote more comments, made fewer structural mistakes, and programmed more systematically.

\subsection{Students' interaction patterns with LLMs}
Amoozadeh et al.~\cite{amoozadeh2024student} studied how students interact with LLMs during programming tasks, identifying three key interaction points (beginning, middle, and after completion) and six problem-solving activities: reading, thinking, writing code, modifying code, prompting, and debugging. They also observed three task decomposition strategies: copying full task descriptions, breaking tasks into subtasks, and a hybrid approach where students solved parts independently while seeking LLM assistance for others. The 60 prompts collected fell into four categories: requesting full solutions, conceptual explanations, program logic clarifications, and debugging help. Building on this, Smith et al.~\cite{smith2024prompting} explored using student responses to Explain in Plain English (EiPE) questions as prompts for code generation with LLMs. Using GPT-3.5 and unit tests, they evaluated how well student-written descriptions translated into correct code, finding this approach effective for teaching precise prompt-writing and improving code comprehension. To assess LLM performance on student-generated prompts, Babe et al.~\cite{babe2023studenteval} developed StudentEval, a benchmark of 1,749 prompts written by novice programmers. Unlike expert-crafted benchmarks, StudentEval captures real-world student interactions, revealing significant variation in prompting techniques and highlighting how LLM nondeterminism can sometimes mislead students about their prompting effectiveness.

\subsection{Developer-LLM conversation patterns}
Hao et al.~\cite{hao2024empirical} manually investigated how developers structure their prompts in multi-turn conversations with LLMs. They identified six patterns in developer-ChatGPT interactions, with five patterns beginning with disclosure of the initial task, followed by iterative follow-up, prompt refinement, requests for clarification, negative feedback, or introduction of a new task. This highlights the dynamic nature of human-LLM interactions in programming contexts. Similarly, Ehsani et al.~\cite{ehsani2025towards} analysed 433 GitHub conversations between developers and ChatGPT, focusing on how prompt knowledge gaps and conversational style affect issue resolution. They found that unresolved issues tend to require more prompts than resolved issues, suggesting that complex issues require longer interactions. The most common conversation styles were directive prompting, chain of thought, and responsive feedback. Notably, prompts with knowledge gaps were more common in open issues (334) than in closed issues (107), with ``Missing Context'' being the most common gap type in both categories. Zamfirescu-Pereira et al.~\cite{zamfirescu2023johnny} provided further insights by examining how non-AI experts approach prompt design when using LLM-based tools. Using a design probe—a prototype chatbot design tool that supports development and systematic evaluation—they found that participants explored prompt design opportunistically rather than systematically and faced challenges similar to those in end-user programming systems. They identified two key barriers to effective prompt design: expectations shaped by human-to-human instructional experiences and a tendency to overgeneralize from limited examples.

\subsection{Prompt engineering strategies for LLM evaluation}
In the area of prompt engineering effectiveness, Kim et al.~\cite{kim2023better} examined how different prompting strategies impact LLM evaluation of natural language generation. They compared Human Guideline (HG) prompts, which mimic human annotation instructions, with Model Guideline (MG) prompts, which provide explicit evaluation steps. Their findings showed that HG prompts aligned more closely with human judgments, while MG prompts, though systematic, were less intuitive. They also found that demonstration examples often introduced bias rather than improving performance and that direct score aggregation correlated best with human evaluations. Expanding on user interactions with LLMs, Srinivasa Ragavan and Alipour~\cite{ragavan2024revisiting} explored how information foraging theory applies to chatbot-based information seeking. They compared traditional web search with chatbot interactions, proposing that users adopt different cost-benefit strategies when engaging with LLMs. Their findings suggest that trust is a key factor shaping how users seek and evaluate information, offering insights for designing more effective LLM-based systems.

\section{{\approach}}
\label{sec:approach}
In this section, we describe how \approach works. We first present the general framework of \approach that can be applied to any dataset, then we discuss the realization of \approach for a dataset.

\subsection{\approach framework}
Prompting in LLM can be understood as a search through a large multidimensional space of potential responses, where each prompt guides the LLM's traversal. Prompts essentially constrain the response state. A successful prompting session will navigate the LLM towards correct response, while a failing prompting fails to do so. \approach takes prompts in a user prompting session and uses few-shot learning to translate each prompt to its corresponding constraints.

\approach uses simple propositional logic formula to capture the semantic representation of the prompts. 
Each formula consists of propositions that denote a statement as well as logical connectives, such as the connective described in \texttt{and} ($\land$), which serve to construct more complex propositions from simpler components.
Basic first-order propositional formulas serve as flexible instruments for expressing various constraints encountered in real-life scenarios. 

In using LLM to solve programming problems, the prompts express constraints that limit the space of responses that LLMs can generate. For example, if a prompt contains ``write a Python function'', it funnels LLM to generate code in the Python programming language and not Javascript or Haskell. The prompts express the requirements of the problem that can further constrain the ways that LLM can generate the code that solves the problem.

Figure~\ref{fig:workflow} depicts the overall workflow in \approach.
\approach uses few-shot learning~\cite{kalluri2022exploring} to learn the structure of prompts in a given dataset and how to translate them into logical constraints. 
To this end, it requires few samples of sequences of prompts in the dataset along with their corresponding constraints (yellow boxes in Figure~\ref{fig:workflow}). In our experiments, only two consecutive prompts from one user interaction sufficed to generate reliable constraints for the entire dataset. 
Then, sequence of prompts that we want to analyze (red box) is appended and send to 
an LLM (purple box) to predict the constraints of each prompt in the sequence. 
We used OpenAI's GPT-4 in our results, which performed sufficiently well. Other LLMs can be used as well.

\begin{figure*}
\begin{tikzpicture}[
    font=\small,
    node distance=0.4cm,
    every node/.style={rounded corners}
]

\node[draw,
      fill=gray!20,
      minimum width=4.5cm,
      minimum height=4cm,
      label=above:{\textbf{A Few Shot Prompt}},
      align=center] (fewshotprompt) {};

\node[draw, fill=yellow!30, text width=4cm, align=center, anchor=north] (ex1) at ($(fewshotprompt.north) + (0,-0.5)$) {\textit{Input:} Prompt 1, \textit{Output:} Constraints in Prompt 1};

\node[draw, fill=yellow!30, text width=4cm, align=center, below=0.2cm of ex1] (ex2) {\textit{Input:} Prompt 2, \textit{Output:} Constraints in Prompt 2};

\node[draw, fill=red!30, text width=4cm, align=center, below=0.2cm of ex2]  (userInput) {\textit{Input:} Prompts under analysis, \textit{Output:} };

\node[draw,
      fill=blue!10,
      text width=2cm,
      align=center,
      minimum height=3cm,
      right=1.8cm of fewshotprompt]
      (model){LLM (GPT-4)};

\node[draw,
      fill=green!20,
      text width=2cm,
      minimum height=2cm,
      align=center,
      right=2.5cm of model] 
      (modelout) {Constraints for the Prompts under analysis};

\draw[->, thick] (model.east) -- (modelout.west);
\draw[->, thick] (fewshotprompt.east) -- (model.west);
\end{tikzpicture}
\caption{Workflow in \approach}
\label{fig:workflow}
\end{figure*}

\subsection{Realization of \approach for programming tasks}
\label{sec:realization}
\begin{figure}
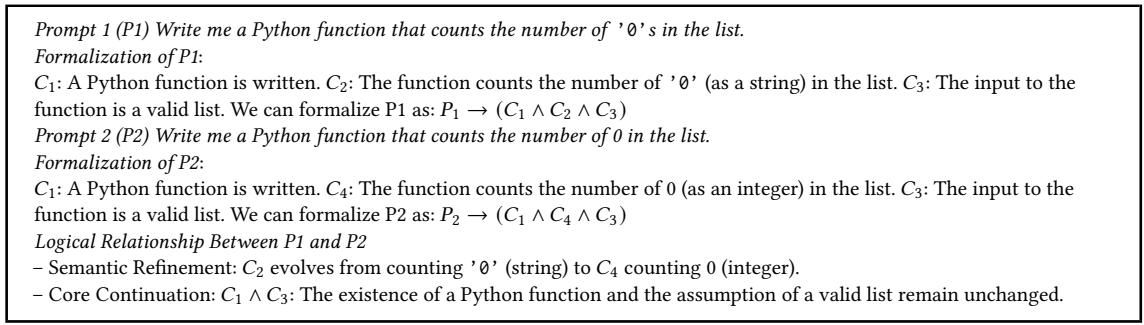

\begin{small}
\begin{mdframed}[linewidth=1pt]
\begin{flushleft}
\textit{Prompt 1 (P1) Write me a Python function that counts the number of \texttt{'0'}s in the list.}

\textit{Formalization of P1}:\\
$C_1$: A Python function is written. $C_2$: The function counts the number of \texttt{'0'} (as a string) in the list. $C_3$: The input to the function is a valid list.
We can formalize P1 as: $P_1 \rightarrow (C_1 \land C_2 \land C_3)$

\textit{Prompt 2 (P2) Write me a Python function that counts the number of 0 in the list.}

\textit{Formalization of P2}: \\
$C_1$: A Python function is written. $C_4$: The function counts the number of 0 (as an integer) in the list. $C_3$: The input to the function is a valid list.
We can formalize P2 as: $P_2 \rightarrow (C_1 \land C_4 \land C_3)$

\textit{Logical Relationship Between P1 and P2}\\
-- {Semantic Refinement}: $C_2$ evolves from counting \texttt{'0'} (string) to $C_4$ counting 0 (integer).\\
-- {Core Continuation}: $C_1 \land C_3$: The existence of a Python function and the assumption of a valid list remain unchanged.
\end{flushleft}
\end{mdframed}
\end{small}
\caption{Few-shot learning examples used in this paper for analysis of the \cite{denny2024prompt}'s dataset.}
\label{fig:few-shot}
\end{figure}

We realize the \approach framework for analyzing students' prompts in programming tasks in a study by Denny et al.~\cite{denny2024prompt}. We thank the authors for sharing their data with us.  This dataset includes students' prompt sequences for three CS1-level programming tasks in Python programming language. 
 Figure~\ref{fig:few-shot} shows the template that we used to instruct LLM how to analyze the prompts. 
This template is structured by the following grammar:

\begin{quote}
Prompt 1 (P1) <text of first prompt> Formalization of P1 <description of individual constraints in P1> We can formalize P1 as: $P_1 \rightarrow$ <logical expression for P1> \\
Prompt 2 (P2) <text of second prompt> Formalization of P2 <description of individual constraints in P2> We can formalize P2 as: $P_2 \rightarrow$ <logical expression for P2> \\
Logical Relationship Between P1 and P2\\
-- Semantic Refinement: <differences in P1 and P2 constraints>\\
-- Core Continuation: <similaities in P1 amd P2 constraints>\\
prompts <new prompts>
formalization:
\end{quote}

In this template, we first provide the LLM,  with an example of two consecutive prompts P1 and P2 (picked from the dataset), 
their corresponding constraints , 
descriptions of each constraint,
and the logical relationship between two prompts.
\approach then instructs the LLM, GPT-4 in this case,  to generate constraints for a new sequence of prompts. We use OpenAI's API to automate the entire workflow. 
GPT-4 returns, description of constraints in each prompt, the propositional logic constraints that represents the prompts, and their relationships.


\subsubsection{Dataset}
\label{sec:dataset}
We use Promptly dataset ~\cite{denny2024prompt} in this paper.
This dataset contains 1872 prompts from 203 CS1 students where they attempted to solve three programming tasks in the Python programming language.
The students were supposed to generate the solution merely by prompting--they were not allowed to edit the LLM's response.

Table~\ref{tab:tasks} 
summarizes these tasks.
The tasks were presented sequentially: only participants who 
successfully completed \textit{counter} were permitted to attempt \textit{initials}, 
and similarly only participants who completed \textit{initials} could proceed to \textit{repeat}.
Participants were also explicitly encouraged to reduce the length of their prompt as much as possible after successfully arriving at the correct prompt. 
The dataset offers insight into how students develop and optimize their prompts.

\begin{table*}[ht]
\caption{Prompt Problems used in the three studies}
\label{tab:tasks}
\centering
\begin{tabular}{|p{0.15\linewidth}|p{0.45\linewidth}|p{0.25\linewidth}|}
\hline
\textbf{Name} & \textbf{Description} & \textbf{Example} \\ \hline
\texttt{1. counter} & Given a list of integers, return how many times the value 0 appears. & \texttt{counter[(10, 20, 30)] => 0} \\ \hline
\texttt{2. initials} & Given a string containing multiple words, return the concatenation of 
the uppercase first letters of each word. & \texttt{initials("abd def ghi") => "ADG"} \\ \hline
\texttt{3. repeat} & Given a list of integers, for each integer \(n\) in the list, 
repeat that integer \(n\) times in the output list. & \texttt{repeat([5]) => [5, 5, 5, 5, 5]} \\ \hline
\end{tabular}
\end{table*}

\subsubsection{Constraint Statistics}
\begin{figure}[H]
    \centering
    \includegraphics[width=0.6\columnwidth]{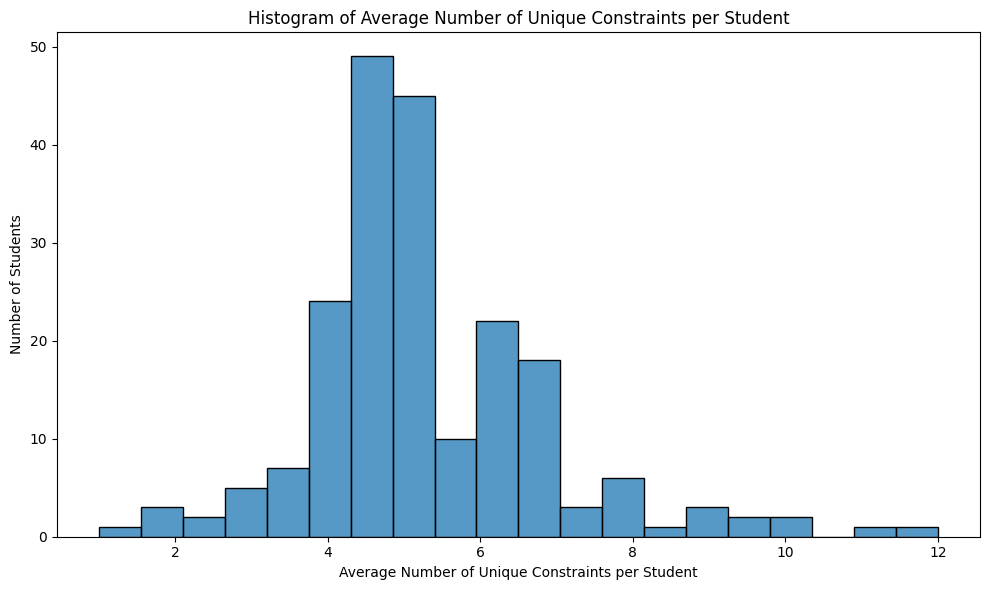}
    \caption{Distribution of Constraints per User}
    \label{fig:histogconstraints}
\end{figure}

Table~\ref{tab:summary} shows the distribution of the number of constraints for each task. Task 1 had the highest average of the number of constraints (5.77) and task 3 had the lowest average number of constraints (4.6). 
Figure~\ref{fig:histogconstraints} shows the distribution of the number of constraints per users. It shows that most students have an average constraint count in the range of about 4 to 5 constraints per task, indicating a moderate level of complexity in their responses. However, there is a noticeable tail of students who use double more constraints (up to around 12), implying that while most students converge on a relatively small to moderate number of constraints, a very small number of students use more complicated constraints in their prompting.

\begin{table*}
\centering
\caption{Summary statistics of the number of constraints in students prompts for each task}
\label{tab:summary}
\begin{tabular}{lccccccccc}
\hline
\textbf{Task} & \textbf{\#Users} & \textbf{Mean} & \textbf{Std} & \textbf{Min} & \textbf{Q1} & \textbf{Median} & \textbf{Q3} & \textbf{Max} \\
\hline
1 & 203 & 5.77 & 2.28 & 1 & 4 & 5 & 7 & 19 \\
2 & 159 & 5.16 & 1.64 & 1 & 4 & 5 & 6 & 11 \\
3 & 146 & 4.60 & 1.32 & 0 & 4 & 4 & 5 & 13 \\
\hline
\end{tabular}
\end{table*}

\section{Evaluation Methodology}
\label{sec:evaluation_methodology}
In this section, we describe the evaluation methodology. To answer the research questions, we evaluate a realization of \approach as described in Section~\ref{sec:realization} as a case study. 

\subsection{RQ1: Measure the accuracy of \approach}
To answer \textbf{RQ1} and measure the accuracy of \approach, we first randomly sample a representative sample of prompts and manually inspect the accuracy of constraints of each prompt. To obtain 95\% confidence interval with 6\% margin of error, we manually randomly sample and inspect 234 prompts.

\subsection{RQ2: Prompt evolution analysis}
To understand how students develop and optimize their prompts, we compare the constraints in consecutive prompts for the changes. 

Since all logical expressions in the prompts were in conjunctive normal form, we could categorize the changes into the following four categories.

\begin{itemize}  
    \item \textbf{Adding Constraints:} A prompt refines the previous prompt if it adds additional constraints to the previous prompt. 
    That is, its constraint is stricter than the previous one.
    \item \textbf{Modifying Constraints:} A prompt revises the previous prompt if it does not contain all the prompts in the previous prompt. It can seem as backtracking in problem solving. 
    \item \textbf{Rewording:} If two consecutive prompts have identical constraints but differ in text, we categorize this as Rewording. It happens when students paraphrase the previous prompt.
    \item \textbf{Resubmission:} The previous prompt without any textual changes is submitted.
\end{itemize}

\begin{table*}
    \centering
    \caption{Evolution Example}
    \begin{tabular}{|c|p{5cm}|c|p{5cm}|} \hline
         Prompt\# & Prompt text  & Translation to constraints& Type of change\\ \hline
        P1 & Write me a Python function that counts how many zeros are in a list of X numbers & $C1 \land C2 \land C3$&  \\ \hline
        P2 & Write me a Python function that counts how many zeros are in a list of X numbers, using the \texttt{.count()} function & $C1 \land C4 \land C2 \land C3$ & Adding a constraint to P1. (A requirement for using \texttt{.count} was added.)\\ \hline
        P3 & Write me a Python function that would respond with \texttt{counter([0, 2, 3, 4, 5, 6, 0]) => 2} when given a list of numbers. This should be counting the number of zeros given. & $C1 \land C5 \land C2 \land C3$& Constraints in P2 have been modified. (The requirement for using \texttt{.count} was removed, and the requirement for function signature was added.)
        \\ \hline
        \end{tabular}
    
    \label{tab:evolution}
\end{table*}

Table~\ref{tab:evolution} shows an example of changes in a sequence of prompts.
Since P1 is the initial prompt, no modification type is assigned.  
P2 introduces a new requirement that the \texttt{.count()} function must be used. Since this adds a stricter condition to the previous prompt without altering other aspects, it is categorized as a adding constraints.
P3 introduces a requirement for a new function name \texttt{counter} with specific output behavior. Furthermore, the requirement to use \texttt{.count()} ($C4$) is removed. Since this change modifies previous conditions rather than simply adding constraints, it is categorized as a modifying constraints.

As an example of the rewording case, \approach generated $C1 \land C2 \land C3 \land C4$ constraints for the following prompt,
\begin{quote}
    ``Write me a python function name counter which contain a list of numbers and return me how many number 0 in the list''
\end{quote}
The user submitted the following prompt,  replacing ``return me how many number 0 in the list'' with ``count(0)''

\begin{quote}
    ``A python function counter with a list of numbers and return count(0)''
\end{quote}

As a result, \approach generated the same constraints as in the previous prompt, since they are semantically equivalent. It shows an interesting case where a user attempts to reduce the length of a prompt by replacing a part of natural language specification with a more succinct equivalent in a formal language.

\subsection{RQ3: Comparison of the difference between successful and unsuccessful prompt sessions}
To understand the difference between successful and unsuccessful prompts, we count the number of changes in constraints between consecutive prompts.
We then use a statistical test to compare the number of changes in students' successful and unsuccessful promptings.

\section{Result}
\label{sec:results}
\subsection{RQ1: Correctness and robustness of the generated constraints}
To evaluate the accuracy of the translation of the prompts to correct logical expressions, we calculated the z score for 95\% confidence interval with a 6\% margin of error for this dataset~\cite{kotrlik2001samplesize}.
We randomly sampled and manually inspected 234 prompts. 
Of 234 samples, \approach translated 225 prompts correctly to propositional logic constraints, and for 9 prompts, the number of constraints generated by \approach was fewer than what an expert would extract. For example, \approach found three constraints for the following prompt:
``Write a Python function that counts the number of zeros in a list and prints it'', that is, $C1$ for a Python function, $C2$ counts the number of zeros, $C3$ prints it. Although plausible, $C2$ contains two equally important constraints for analysis (1) counting and (2) zero values. Therefore, we expected \approach to generate four constraints, not three. 

During our manual inspection, we found that \approach and GPT-4 are remarkably robust to reordering the text in the prompts. For example, it outputs $C1 \land C4 \land C2 \land C5 \land C3$ for 
    ``Write me a Python function named counter. The counter function takes a list as its input and returns the number of zeros in that list.''
It also generates an equivalent logical expression $C1 \land C2 \land C5 \land C3 \land C4$, for a subsequent prompt that preserves the correspondence of constraints with reordered text:
``Write me a Python function that counts the number of 0
     in a list of integers and returns the value as its output.
     The function is called counter.''

\subsection{RQ2: Prompt evolution analysis}

\begin{figure}
    \centering
    \begin{subfigure}[b]{0.7\textwidth}
    \includegraphics[width=0.95\textwidth]{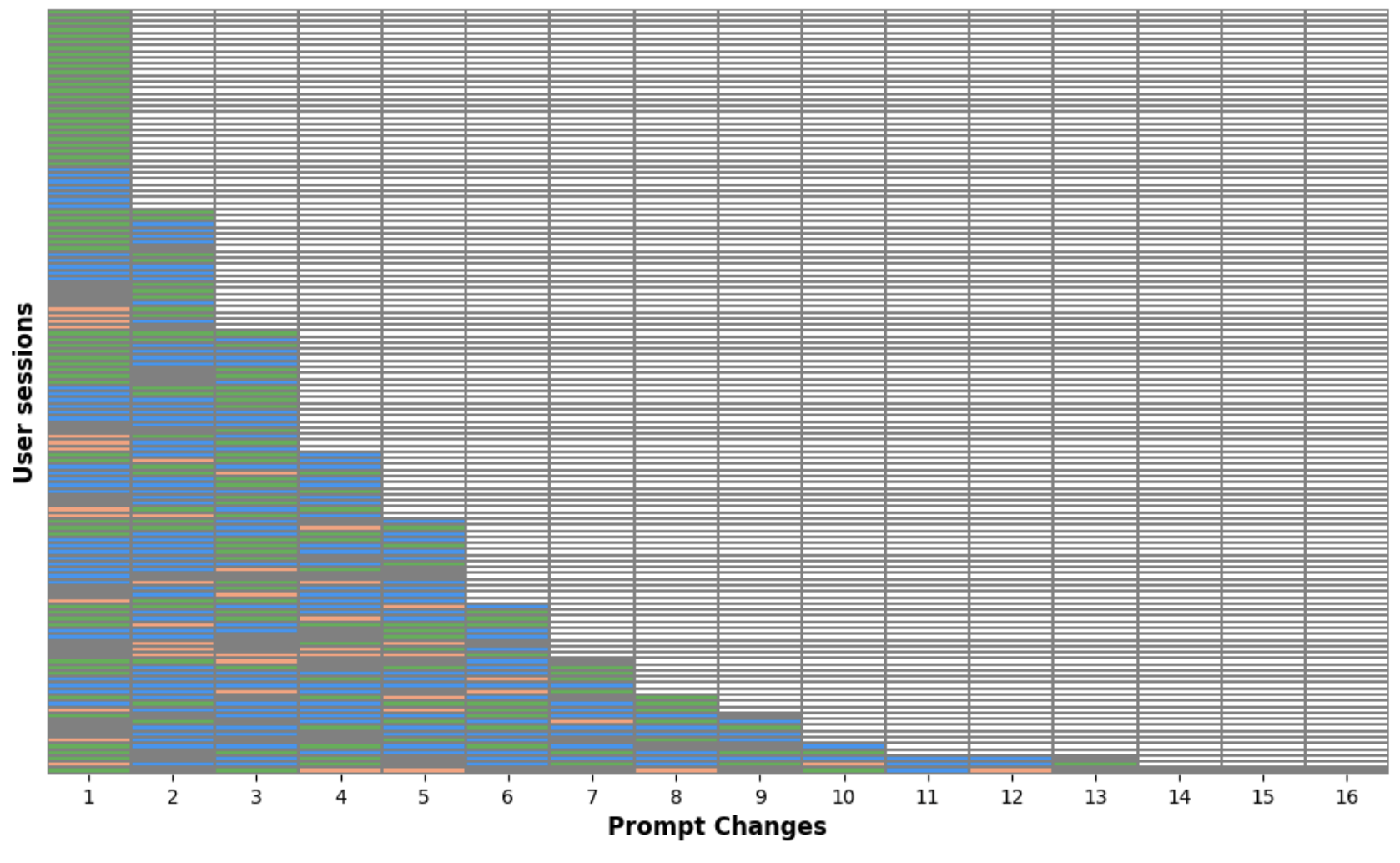}    
    \caption{Task 1}
    \label{subfig:task1}
    \end{subfigure}
        \begin{subfigure}[b]{0.7\textwidth}
    \includegraphics[width=0.95\textwidth]{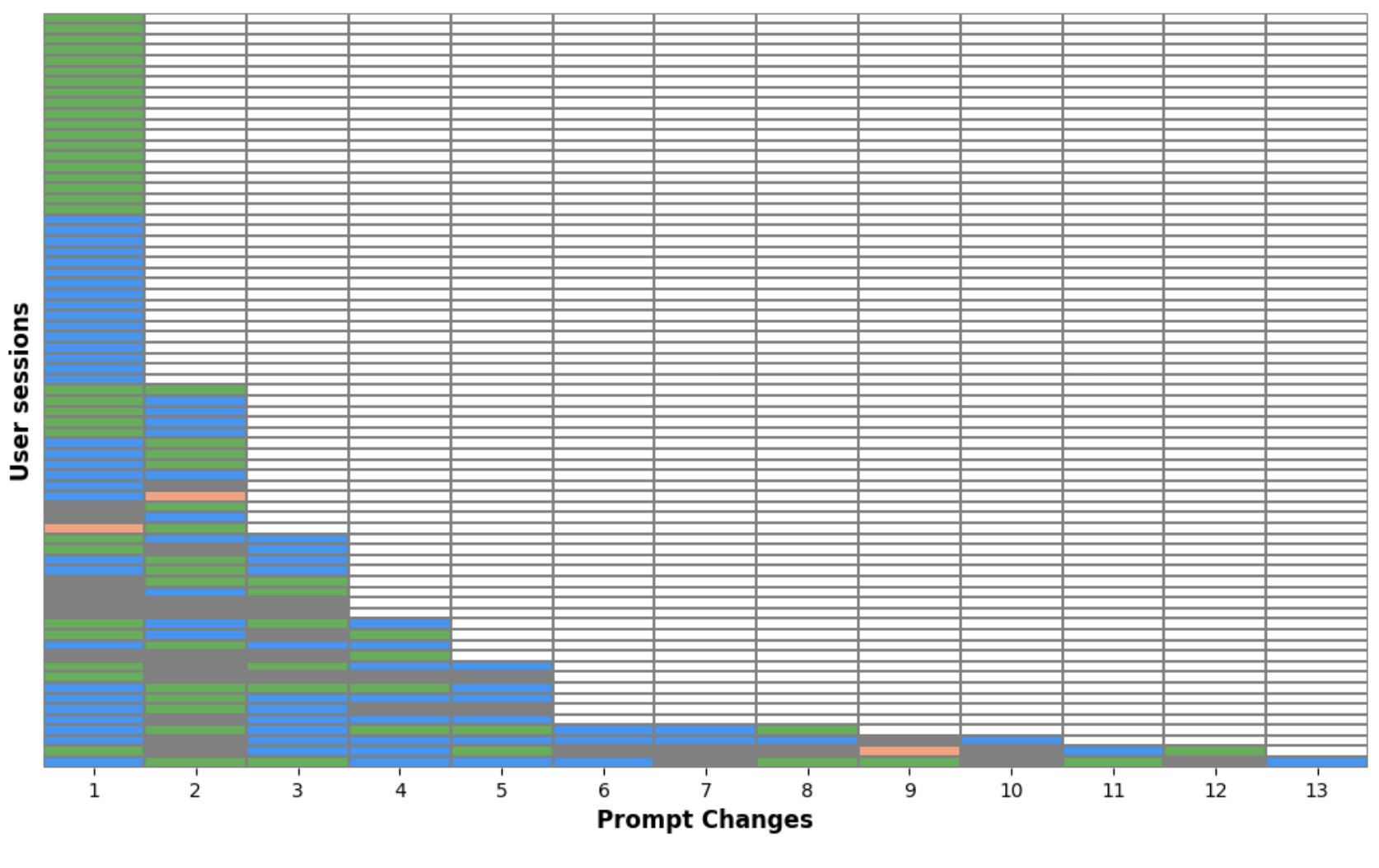}    
    \caption{Task 2}
    \label{subfig:task2}
    \end{subfigure}
\begin{subfigure}[b]{0.7\textwidth}
    \includegraphics[width=0.95\textwidth]{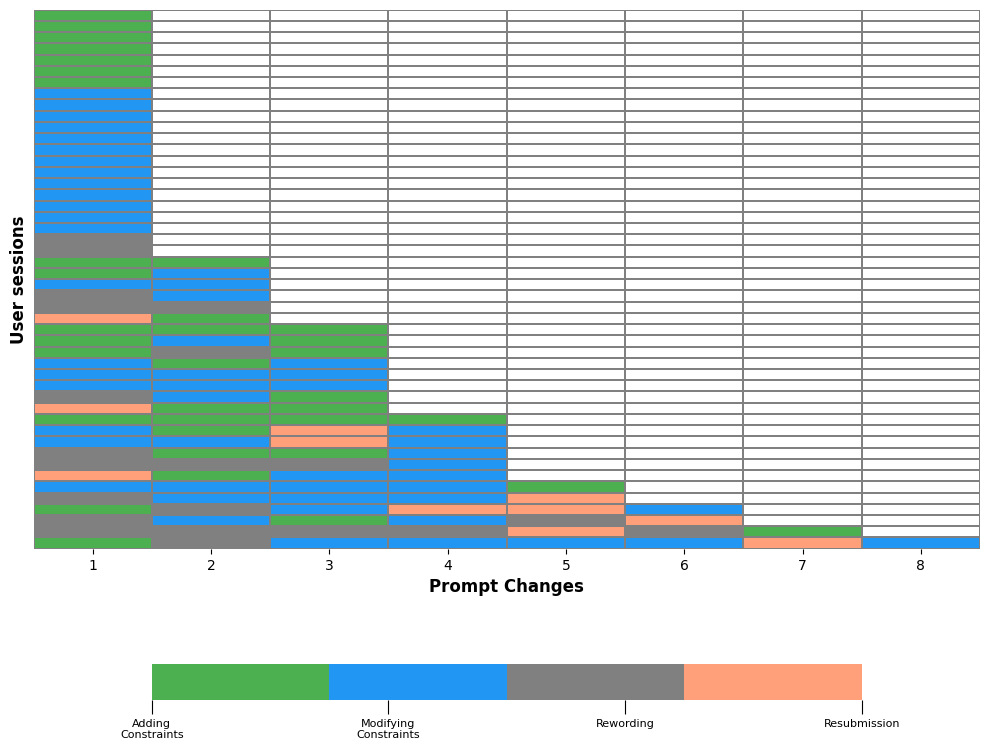}    
    \caption{Task 3}
    \label{subfig:task3}
    \end{subfigure}
    \caption{Changes in the constraints for each prompting task for students who finished all three tasks}
    \label{fig:heatmaps}
\end{figure}

Figure ~\ref{fig:heatmaps} shows the types of changes in the successful prompt sequences of the students. First, the student starts with an initial prompt. Second, if the student does not get the right solution from the first prompt, they will begin adding more requirements (adding constraints), or modifying existing requirements in the prompt (modifying constraints). Additionally, some students resubmit the same prompt (resubmission) or rephrase it (rewording). 
The figure shows that the students prompted more in the first task compared to the subsequent tasks.  We investigated a potential correlation between the type of changes, and the length of prompting sessions.

\subsubsection{Correlation of Constraint Changes and the Length of Prompting}
Table \ref{tab:corr} presents the results of the Pearson correlation analysis between percentages of the number of constraint changes in each sequence of prompts with the length of the sequences. It suggests a moderate negative correlation between the percentage of adding constraints with the length of prompting, which means that students who frequently add constraints, their prompting sequence tends to be shorter. This may suggest that adding constraints helps students converge on a solution faster and can be due to their core refining of their prompts. On the other hand, weak positive correlation between the length of prompting and the other constraint changes: modifying constraints, rewording, and resubmission, which may suggest that these constraints changes slightly increase the prompting sequence. This can be an indication of trail and error where students try to remove part of the prompt and introduce a new part in modifying constraints, trying rewording the same prompt, or just resubmit the same prompt. 

\begin{table}
\centering
\caption{Results of Pearson correlation between changes in the constraints and problem solving length}
\begin{tabular}{lcc}
\hline
\textbf{Activity} & \textbf{Correlation} & \textbf{p-value} \\
\hline
Adding Constraints & -0.338464 & 0.000004 \\
Modifying Constraints & 0.177237 & 0.018612 \\
Rewording & 0.181198 & 0.016098 \\
Resubmission & 0.157528 & 0.036797 \\
\hline
\label{tab:corr}
\end{tabular}
\end{table}

\subsubsection{\approach for Understanding Problem Solving Steps}

\begin{figure}
\begin{subfigure}[b]{0.45\textwidth}
    \centering
    \includegraphics[width=0.8\columnwidth]{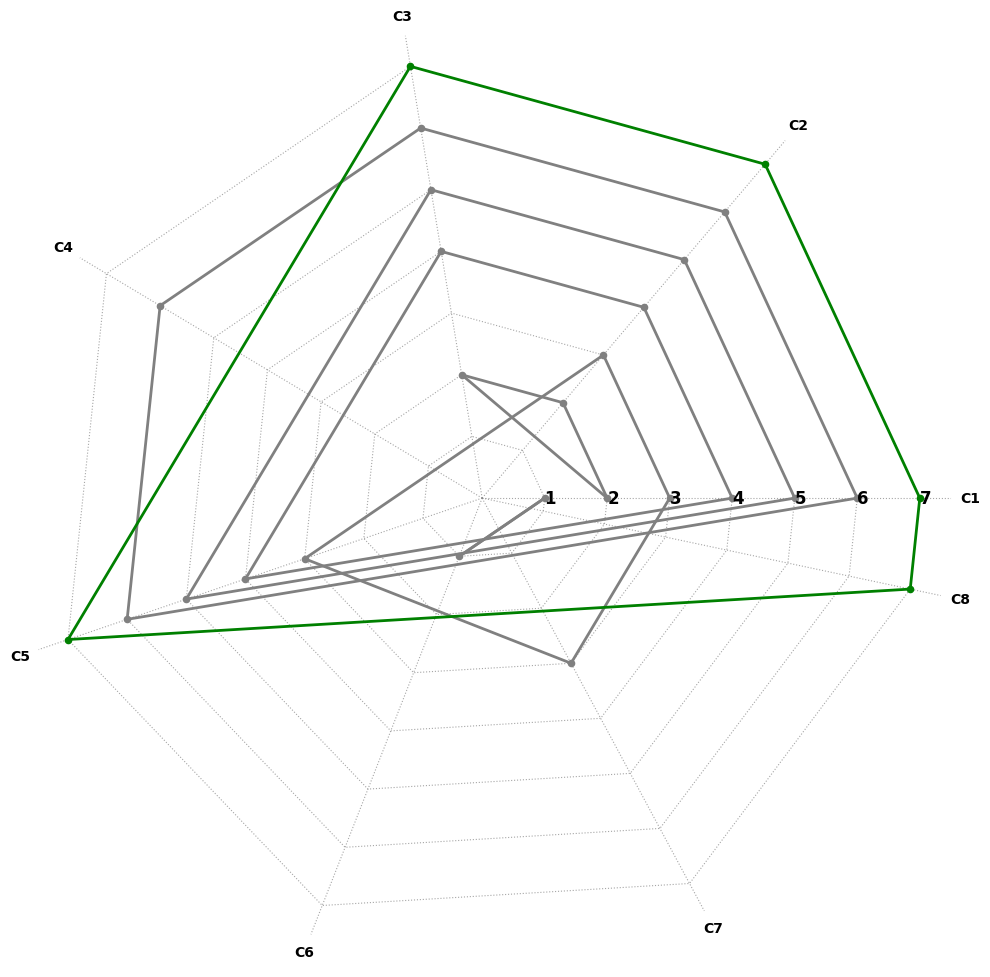}
    \caption{Successful}
    \label{fig:SucessScenario}
\end{subfigure}
 \hfill
\begin{subfigure}[b]{0.45\textwidth}
    \centering
    \includegraphics[width=0.8\columnwidth]{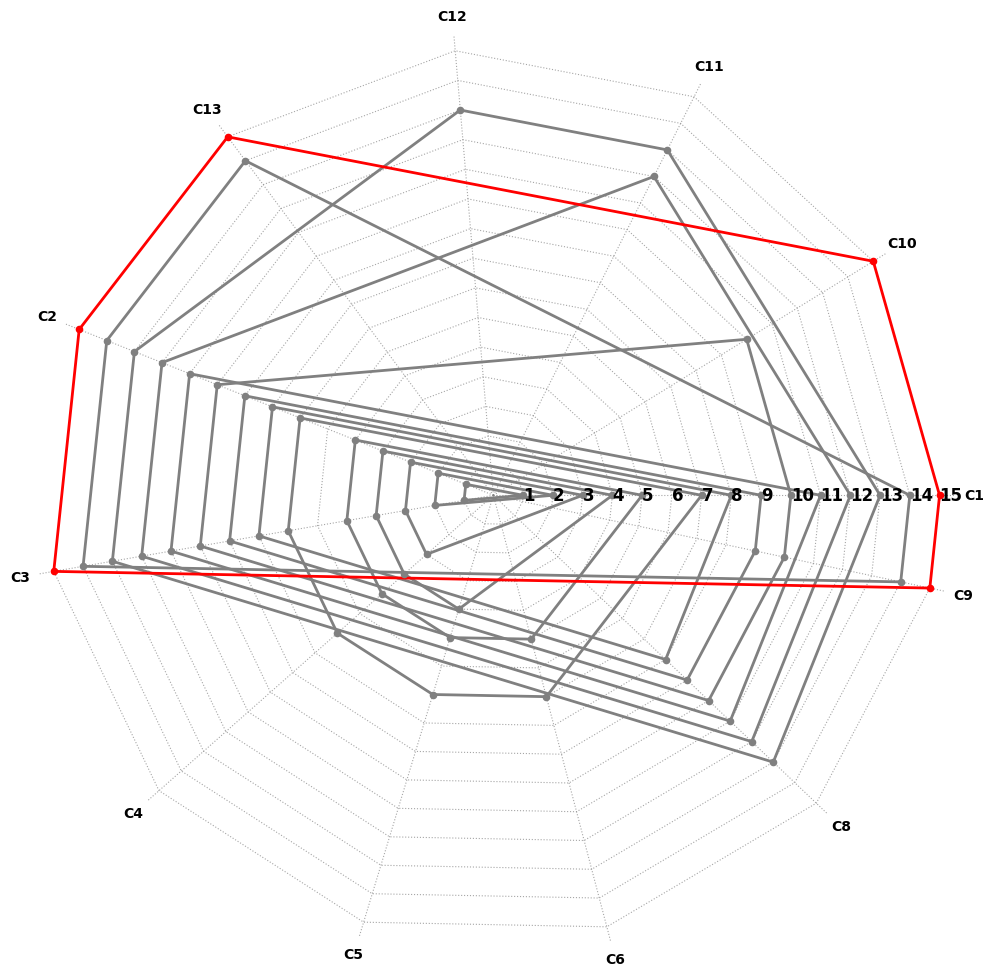}
    \caption{Failed Prompting}
    \label{fig:FaieldScenario}
\end{subfigure}

    \caption{Changes in constraints}
    \label{fig:spider:changes}
\end{figure}

Using \approach, the constraints in each prompt can provide insight into how students try to guide the LLM to the solution. In this section, we demonstrate how the constraints can be used to analyze student prompting strategies at the level of constraints and how they could potentially be used to find where an intervention to help students can be helpful. We note that we will use two examples to motivate the potential solution. In future work, a thorough evaluation of the generalizability and usefulness of the proposed approach in pedagogy will still be needed.

Figure~\ref{fig:spider:changes} shows examples of changes in the constraints of two users in problem 1, Figures~\ref{fig:SucessScenario} and~\ref{fig:FaieldScenario} show a successful and a failed attempt to solve the problem successfully, respectively. Each axis in the spider graph represents a constraint, and scales represent prompt orders where the innermost scale represents the first prompt and the outermost scale represents the last prompt. 
Figure~\ref{fig:SucessScenario} shows that the student starts with a couple of constraints initially. The constraints in the second and third prompts suggest that the student evaluates different approaches, by trying $C3$ in the second prompt, then replacing it in the third prompt with $C5\land C7$. From the fourth prompt, the student picks pieces from the last two prompts $C3\land C5$ and incrementally adds more constraints in prompts 5 and 6. In the last prompt, the last prompt is slightly modified (removing $C4$) and adding $C8$. Overall, except for initial prompts, the changes in the constraint are limited to a limited number of constraints. 

In contrast, Figure ~\ref{fig:FaieldScenario} shows an example of a failed attempt to solve problems. Unlike the previous problem solving approach, we see that while the student incrementally adds more constraints to the initial prompts until prompt 8, there is a sudden large change in the number of prompts in prompt 6 that suggests that the student may have changed their solution approach. Moreover, the rate of change in the subsequent prompt increases substantially, with two or three constraints that may suggest the student struggle in formulating a solution. 

\paragraph{Opportunity} The sudden changes in the constraints in Figure ~\ref{fig:FaieldScenario} may suggest students assumes that they're exhausted the current approach and a drastic change in the problem solving or prompting strategy.
It may suggest that such points can be good candidates for an intervention, e.g., generating hints. For example, manual inspection of the prompts in Figure ~\ref{fig:FaieldScenario} showed that the student was very close to the solution in their third prompt, and only misplaced the name of the function. As the frustration of the students grows, their sixth prompt becomes: ``it is stupid'', and afterward the user explores other parts of the solution space. Therefore, \approach can suggest the appropriate times for an intervention.

\subsubsection{Understanding Prompt Length Reduction with Constraints}

In the prompt problem study~\cite{denny2024prompt}, students were encouraged to minimize their successful prompts.
Applying \approach would allow us to see if the students made substantial changes to the semantics of the prompts or merely tried to reduce the length of the prompts by rewording. These two approaches would entail two different approaches. In the first one, the student would need to rethink the constraints in the solution, while the second approach may mostly focus on wordsmithing.

Table~\ref{tab:reducedprompts} shows the results.
We identified the first successful prompt and the subsequent smaller successful prompts and compared their constraints.
Out of 41 pairs, we found that 35 smaller prompts have the same constraints as the original correct prompts, five reduced prompts had fewer constraints than the original prompts, and only one had more constraints than the original prompts. It suggests that most of the students resorted to rewording and wordsmithing to reduce the length of the prompts.

\begin{table}

    \centering
    \caption{Relation between constraints in the reduced prompts and original prompts}
    \begin{tabular}{|c|c|c|c|} \hline
       Relation  & Identical constraints & Less constraints & More constraints \\ \hline
       Number & 35 & 5 & 1\\ \hline
    \end{tabular}
    
    \label{tab:reducedprompts}
\end{table}

During the analysis, we found an interesting case in which a significant reduction in word count did not affect the number of identified constraints. 
Specifically, some users used underscores to combine multiple words into a single token, reducing the overall word count to just \textbf{one word} without altering the prompt's content, task requirements, or constraints.

\subsubsection{Number of constraints}
\begin{figure}
\begin{subfigure}[b]{0.45\textwidth}
    \centering
    \includegraphics[width=0.95\columnwidth]{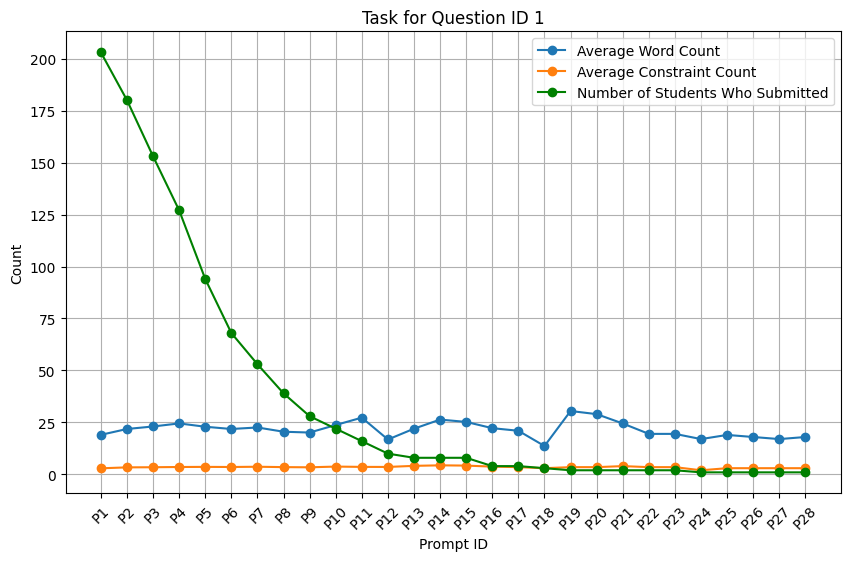}
    \caption{Task 1}
    \label{fig:NoOfCoQ1}
\end{subfigure}
\hfill
\begin{subfigure}[b]{0.45\textwidth}
    \centering
    \includegraphics[width=0.95\columnwidth]{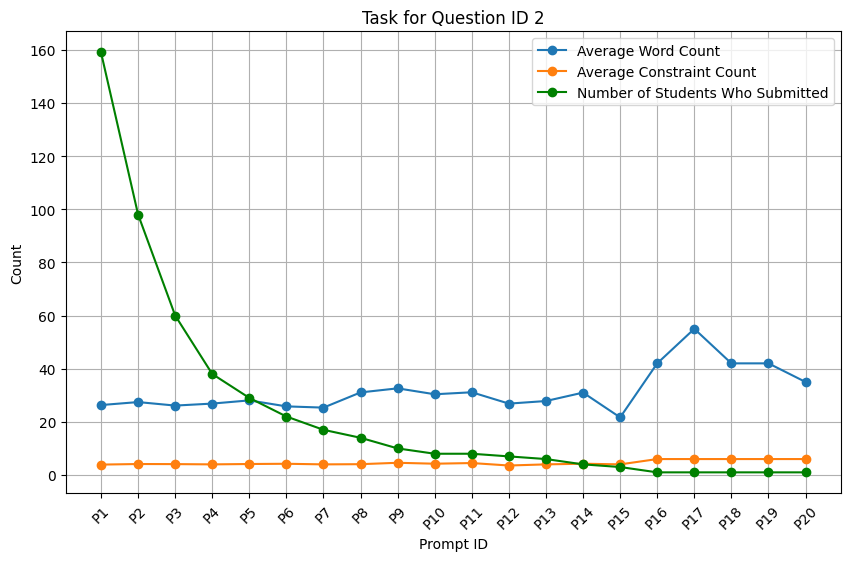}
    \caption{Task 2}
    \label{fig:NoOfCoQ2}
\end{subfigure}
\hfill
\begin{subfigure}[b]{0.45\textwidth}
    \centering
    \includegraphics[width=0.95\columnwidth]{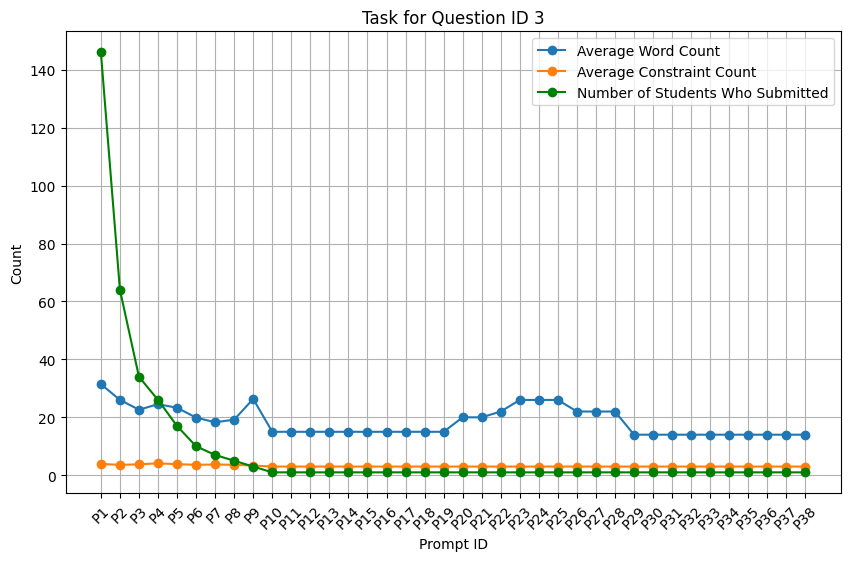}
    \caption{Task 3}
    \label{fig:NoOfCoQ3}
\end{subfigure}
\caption{The average number of words in each subsequent submission compared to the average number of  constraints and the number of participants that submitted}
\label{fig:promptlyresults}
\end{figure}

Figure~\ref{fig:promptlyresults} shows the average number of words (blue line), the number of users (green line), and the average number of constraints (orange line) in each prompt step. 
In the original prompt problem study only the first two were reported, which would only provide a partial view of the textual representation of the prompts and the solutions that they expressed. \approach can enrich it by adding a sematic view of the prompts using constraints. For example, while the figures show some increases in the length of prompts in the long tail of the graphs, the average number of prompts remain relatively steady.

Furthermore, we can see in Figure \ref{fig:NoOfCoQ3} that a student continues to submit 38 prompts and the number of constraints and words is almost the same in most submissions. When we look back at the sequence of strategies in Figure \ref{fig:heatmaps}, we found that most of the time this student was submitting the same prompt without any changes in the language and constraints. 

\subsection{RQ3: Comparing size of constraint changes in successful and unsuccessful prompt sequences}
\begin{figure}
    \centering
    \includegraphics[width=0.95\columnwidth]{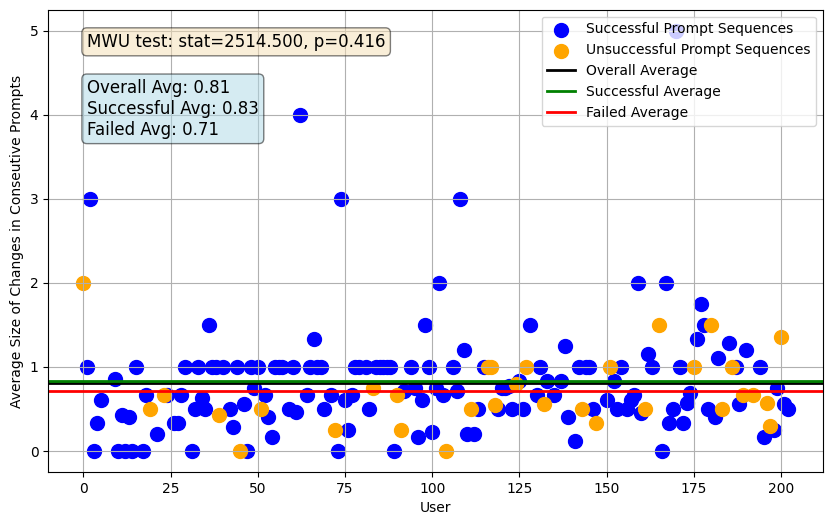}
    \caption{Size of changes in consecutive prompts}
    \label{fig:CoDv2}
\end{figure}

Figure \ref{fig:CoDv2} shows the size of the changes between constraints in consecutive prompts, the green line represents the average size of the differences in successful prompt sequences (0.83), while the red line represents the average size of the differences in unsuccessful prompt sequences (0.71).
The black line denotes the average size of differences in all prompt sequences (0.81). Mann-Whitney U-test failed to find a statistically significant difference between the size of changes in consecutive prompts in 
in the successful and unsuccessful prompt sequences (p-value = 0.416). 

\section{\approach for Measuring the Progress}
In this section, we describe a potential application of \approach in the support of intelligent tutoring systems in the era of generative AI and LLMs. In particular, we show that the constraints can be applied to measure how far or close students are from the correct answer. We outline an algorithm and motivate it in a working example. 

\subsection{Algorithm}
Suppose $S$ is a successful prompt and a sequence $P$ of prompts written by students/users. The algorithm~\ref{alg} outlines an algorithm for measuring the progress of the students toward a solution. The measure can potentially be used to direct an intelligent tutor system or hint generator. 

First, we create a new set $T$ that contains $S$ the prompts in $P$.
Then, we use \approach as previously described to extract the constraints of the prompts $C$.
In the next step, the difference ($diff$) between the constraints of $S$ ($C[S]$) and the constraints of the user prompts are calculated. The result of $diff$ function can be used to measure how far is the students' prompts to the solutions. We note that many different prompts can be a solution to a programming task, in such cases, all representative solution prompts can be collected in a set $S$, and the algorithm will need to compute the difference between the prompts and each of the representative solutions.

\begin{algorithm}
\caption{An algorithm for measuring the success}\label{alg}
\begin{algorithmic}
    \State $T \gets P \cup S$
    \State $C \gets \approach(T)$
    \For{$p \in (C \setminus S)$}
        \State $D[p] \gets \mathrm{diff}(C[S], C[p])$
    \EndFor
\end{algorithmic}
\end{algorithm}

\begin{figure}
    \centering
    \includegraphics[width=0.45\columnwidth]{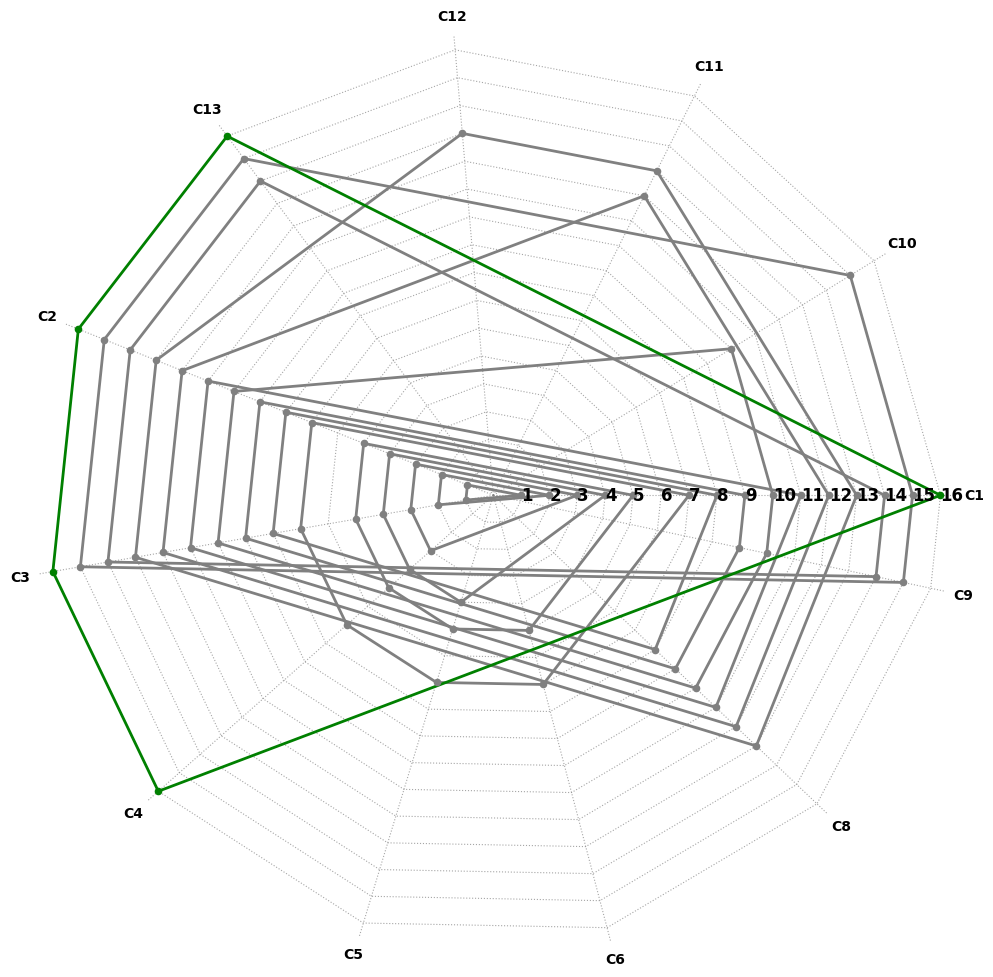}
    \caption{Using constraints to guide failing prompting}
    \label{fig:FaieldScenarioPass}
\end{figure}

\paragraph{A Case Study}
Here we demonstrate the application of ~\ref{alg} to failed prompting. 
We use the failed prompts in Figure ~\ref{fig:FaieldScenario} along with the last prompt in Figure~\ref{fig:SucessScenario} which is a successful prompt. Figure~\ref{fig:FaieldScenarioPass} depicts the resulting constraints for the failed prompts (prompts 1 to 15) along with constraints in the successful prompt (prompt 16--green lines).
Note that the constraints' names changed because we added prompt 16. 

This student was very close to the solution in the third prompt (P3) as it is only one constraint different from the proposed solution in P16. The text for P3 is ``Write me a Python function that uses the counter function and counts the number of elements in the list which have the value '0'.''
After P3 the distance between the student's solution and the proposed solution increases, which leads to changing the direction in problem solution approach from P7. 
The text for P15 is ``you are given a list in python. use the for in range loop to traverse this list. within that list, if there is a 0, add 1 to a variable named counter. once you are done traversing the list, print the variable which is storing the number of 0s found'' and the corresponding constraints that \approach suggests are $1 \land C2 \land C3 \land C9 \land C10 \land C13$.
P16 is ``Write me a Python function that defines the function 'counter' so the output finds and print the number of times an object in the list equals 0'' with the corresponding constraints $C1 \land C4 \land C2 \land C3 \land C13$.
Using the built-in explanation for the logical differences of two prompts, \approach explains ``The functionality remains consistent with P15, without the specific loop and variable requirements.'' 
A future hint generator can either encourage the student to continue refining P3.

\section{Threats to Validity}
\label{sec:threats}

An important limitation of our work is the extent to which findings generalize across different tasks, domains, and learner populations. Because we evaluated \approach using prompts from introductory Python problems, its performance may vary when applied to more advanced programming tasks or to subjects outside computing. Second, the approach relies heavily on LLMs—particularly GPT-4 --for its constraint extraction, which introduces potential bias or instability due to the proprietary and continually evolving nature of these models. Changes in LLM behavior or differences between models could lead to inconsistent results over time. Third, any errors in the few-shot learning examples or prompts used to train \approach can propagate through the system, impacting accuracy in ways that are difficult to detect without additional human review. Finally, like most automated methods, \approach may fail to capture nuanced or context-specific information not readily translatable into propositional constraints. These factors could limit the technique’s ability to fully capture the diverse ways learners conceptualize and express problem requirements. Careful consideration of these threats, and additional validation efforts, such as manual annotation or triangulation with other data sources, are critical for ensuring robust and trustworthy results.

\section{Discussion}
\label{sec:discussion}
In this section, we discuss the results and potential applications presented in the paper. 

\paragraph{Accelerating research}
In our experiments, the GPT-4 few-shot learning performed very well in identifying the constraints with only two examples from the dataset. 
A major benefit of \approach lies in its potential to rapidly scale analyses that would otherwise require exhaustive manual annotation. By automatically extracting logical constraints from student prompts, researchers can efficiently sift through large datasets, pinpointing trends or anomalies with minimal human intervention. This automation not only saves considerable time, but also reduces potential human errors or inconsistencies in classification. Moreover, because it employs few-shot learning, \approach can quickly adapt to new tasks or domains, further broadening its applicability. As a result, researchers can explore hypotheses more easily, conduct comparative studies, and refine their experiments in an iterative manner, ultimately accelerating progress and innovation in both computing education and broader LLM-based investigations.

\paragraph{Finding when students struggle}
\approach provides a way to systematically track a student’s evolving prompt constraints over time, revealing moments where significant or erratic shifts occur. These abrupt changes in constraints—such as suddenly discarding previously used requirements or drastically revising the approach—often signal confusion or frustration, suggesting the student has reached an impasse. By mapping each prompt to propositional logic, educators or automated tutoring systems can quantify the “distance” between a student’s current prompt and a known solution or optimal set of constraints. In practice, this enables real-time detection of where students stray too far from an effective strategy or repeatedly circle back without making meaningful progress. Consequently, instructors can provide targeted interventions, such as clarifying conceptual misunderstandings or offering smaller, more manageable sub-goals. This proactive assistance reduces students’ trial-and-error cycles, leading to a more efficient and supportive learning process.

\paragraph{Potential use in hint generation}
Another advantage of the \approach framework is that it can be integrated into real-time hint generation. By mapping student prompts to logical constraints, the system can measure how closely these constraints align with, or diverge from, a known correct solution’s constraints. This difference can directly inform targeted hints: for instance, identifying when a necessary condition--such as checking list bounds or returning the right data type--is missing, or when two constraints conflict. Moreover, by tracking how a student’s constraints change between prompts, \approach helps pinpoint exactly where they might need support, allowing the system to offer context-aware nudges that guide them back on course. These timely, tailored hints can ultimately help learners refine their problem-solving strategies, reduce trial-and-error cycles, and deepen conceptual understanding.

\section{Conclusion}
\label{sec:conclusion}
In this work, we introduced the \approach approach to systematically analyze students' prompting behavior in programming tasks. The approach is based on the few-shot prompting LLMs to extract constrains that users have specified in the prompt. Our results show that the approach is robust, and provides a multitude of different options for analyzing prompting behavior. Moreover, struggling students tend to modify their prompts more suggesting shifts in the problem-solving strategies. 

This work contributes to scalable prompts analysis which we used to analyze more than 1500 prompts of students interaction with LLM to generate codes for programming tasks. The \approach has potential in real-time LLM educational tools to guide students throughout the prompting journey, and trigger the need for assistance for struggling students when they shift to explore another area than the solution space. In future research, we could extend the \approach to more complex programming tasks.

\bibliographystyle{acm}
\bibliography{references,amin}
\end{document}